\documentclass[prc,aps,twocolumn,showpacs,floatfix,nofootinbib,bibnotes]{revtex4}
\usepackage{epsf}
\newcommand{\ds}{\displaystyle}
\newcommand{\no}{\nonumber}
\newcommand{\dsf}{\ds\frac}
\newcommand{\Tr}{\mbox{Tr}}
\newcommand{\re}[1]{(\ref{#1})}
\begin{document}
\sloppy
%%%%%%%%%%%%%%%%%%%%%%%%%%%%%%%%%%

\title{Nucleon-nucleon potential in finite nuclei}

\author{U.T. Yakhshiev$^{1,2}$\thanks{Electronic address:~u.yakhshiev@nuuz.uzsci.net},
Ulf-G. Mei{\ss}ner$^{3,4}$\thanks{Electronic address:~meissner@itkp.uni-bonn.de},
A. Wirzba$^{3,4}$\thanks{Electronic address:~a.wirzba@fz-juelich.de},
A.M. Rakhimov$^{5}$, M.M. Musakhanov$^{2}$}

\affiliation{
${}^1$
Department of Physics and Nuclear Physics \& Radiation Technology Institute
(NuRI),
Pusan National University, 609-735 Busan, Republic of Korea\\
${}^2$
Theoretical Physics Department,
National University of Uzbekistan, Tashkent-174, Uzbekistan\\
${}^3$
Helmholtz Institut f\"ur Strahlen- und Kernphysik (Theorie),
Universit\"at Bonn, Nu{\ss}allee 14-16, D-53115 Bonn, Germany\\
${}^4$
Forschungszentrum J{\" u}lich, Institut f{\" u}r Kernphysik
(Theorie),  D-52425  J{\" u}lich, Germany\\
${}^5$
Institute of Nuclear Physics, Academy of Sciences, Uzbekistan
}

\date{\today}

\begin{abstract}
We consider the spin-isospin--independent central
part of the residual nucleon-nucleon potential in finite spherical nuclei
taking into account the deformation effects of the nucleons within
the surrounding nuclear environment. It is shown that inside the
nucleus the short-range repulsive contribution of the potential is increased
and the intermediate attraction is decreased. We identify the 
growth of the radial component of the 
spin-isospin independent short-range part of the in-medium  nucleon-nucleon
interaction as the responsible agent that
prevents the radial collapse of the nucleus.
\end{abstract}

 
\maketitle

%\keywords{
%Keywords:
%Skyrme model, nuclear medium,
%soliton deformation, nucleon quadrupole moment.
%}

\section{Introduction}
\label{intro}

Nearly 20 years ago, in their pioneering work~\cite{jjp}, Jackson, Jackson and
Pasquier investigated the skyrmion--skyrmion interaction 
by considering simple
deformations of the chiral field. This idea was further developed in
Refs.~\cite{Kaulfuss,Hajduk,Otofujii} into 
various versions of Skyrme-like
models allowing for more complicated deformations. It was shown that 
the central potential of the original Skyrme model 
is reduced by an amount of about
$10 \div 20$\,\% if deformation effects of 
the skyrmion are taken into account. 
In the $\omega$--stabilized Skyrme model the deformation effect is 
even more significant, 
as  the central repulsion at $R$ = 1\,fm acquires a reduction 
of about 40~\%.

Nowadays, in-medium properties of the nucleon are of high relevance.
%citation
In particular,  properties of a single nucleon embedded in various finite
nuclei have been considered in a medium-modified
version of the Skyrme model~\cite{npa002,epj001}.
It has been shown that bound nucleons acquire an intrinsic
quadrupole moment due to the in-medium deformation
effects. In summary,
the general behaviour
of the properties of the nucleons are in qualitatively
agreement with experimental
observations, e.g.\ the swelling of the nucleon in the
nuclear medium and the decrease of
its mass. Similar results have also been
obtained in the infinite nuclear matter
approach of Ref.\,\cite{prc001}.

Thus, it is only natural to investigate the nucleon--nucleon
(NN) interactions in the nuclear medium because of their role in 
the formation of nuclear matter. 
There are two known paths to nuclear matter aspects 
of the Skyrme model.
The first one is based on  studies of the crystalline ground state of the
skyrmionic
matter~\cite{Klebanov85,Glendenning86,Wust87,Goldhaber87,Kugler,Jackson88,Castillejo89,Forkel89,Baskerville96},
while the second one is related to studies of Skyrme model on the
hypersphere~\cite{Manton86,Manton87,Jackson,JaWiMa,WiBang,Jackson91,Wirzba}.  
Especially in the first
approach, the tensor part
of the potential is found to be responsible for
the crystalline structure of the ground state. 
In contrast to this, quantum
hadrodynamics studies~\cite{Serot86} assert the important role of the
spin-isospin independent scalar part of the NN interactions in the 
formation of
the nuclear matter. In any case, one can expect that 
the surrounding nuclear environment will leave its mark on the
in-medium NN potential.

The temperature and density dependence of nucleon--nucleon interactions,
hadron properties and meson--nucleon coupling constants of 
the one--boson--exchange potential have been studied 
in the framework of thermo field dynamics (TFD) in 
Refs.~\cite{sgao,Zhang97,Alvarez-Ruso96,Rakhimov99,Yakhshiev03}.
In particular, it has been shown that the potential well of the NN
interaction becomes shallow as the temperature or density 
increases~\cite{sgao}. The changes in the meson--nucleon couplings and 
hadron properties also indicate the influence of the medium 
on the one boson exchange 
potential~\cite{Rakhimov99,Yakhshiev03}.
Note, however, that the Skyrme model and its variants are the only type of
models that use hadronic degrees of freedom and that allow for a simultaneous
description of both single-baryon and multi-baryon properties. In fact, as
discussed in Ref.\,\cite{npa002}, Skyrme-type models even allow for a simple
inclusion of nuclear background terms.

The alterations of the tensor part of the 
NN potential for {\em infinite} nuclear matter
have already been studied in the framework of the in-medium modified Skyrme 
model in Ref.\,\cite{prc001}.  In the present paper, we rather concentrate 
on the 
spin-isospin--independent part of the NN potential in {\em finite}
nuclei 
where we allow the chiral fields of the two involved nucleons  to deform 
in  response to the surrounding  nuclear environment.

For simplicity, we will only consider 
the residual in-medium NN potential in the
{\em radial} direction of the nucleus (i.e.\ the {\em axial symmetric case}), 
such that
the two nucleons are subject to
different densities. This should be contrasted with the case where the
in-medium  NN interaction is considered in the azimuthal direction, namely 
where
both nucleons are at the same distance from the center of the nucleus and
therefore subject to  the same density.
This is not a serious  limitation since the 
latter interaction ought to be subdominant to the interaction in the 
radial direction,
as the medium modifications due to the changes in the profile
functions are known to be small, whereas the changes resulting from the
density dependence of the mass functionals are more 
pronounced\,\cite{npa002}.

\section{The model}
Unfortunately, the NN potential of the original Skyrme model does not have a  
central attraction at intermediate distances.
However  it has been shown that this long standing problem can be solved
e.g.\ by the incorporation of the iterated two-pion 
exchange~\cite{NobiUlf1,NobiUlf2} into the Skyrme model or 
by a reformulation~\cite{Rakhimov} of the original Skyrme
lagrangian on the basis of scale invariance and 
the conformal anomaly~\cite{Gomm,Andrianov88}.
Such a scale-invariant lagrangian was also  used  to describe
nucleons in the nuclear matter~\cite{Brown1,Brown2} in the mean-field
approximation (MFA).

\subsection{Lagrangian and baryon number one soliton}
We start here with the model of Refs.\,\cite{Musakhanov} that proposes 
a scale-invariant version of the in-medium Skyrme
model given, in  the {\em static} case,  by the following
lagrangian:
\begin{eqnarray}
{\cal L}_{\rm st}(U,\sigma)&=& 
-\dsf{F_\pi^{2}}{16}\left[\chi^{2}\alpha_p ({\vec x})
{\Tr}(\vec\nabla_{\vec x}{U})(\vec\nabla_{\vec x}{U^+}\!)
\!+\!2(\vec\nabla_{\vec x} \chi)^{2}\right]\nonumber\\
&&\mbox{}+\dsf{1}{32e^{2}}{\Tr}\left[U^+\dsf{\partial}{\partial x_i} U,
U^+\dsf{\partial}{\partial x_j}U\right]^2 \nonumber\\
&&\mbox{}-\dsf{C_g^*}{24}\left[\chi^4 -1
+\dsf{4}{\varepsilon} (1-\chi^\varepsilon)\right]\nonumber\\ 
&&\mbox{}-
\dsf{F_\pi^{2}m_\pi^{2}}{16}\chi^3 \alpha_s ({\vec x})\,
\Tr\left(2-U-U^{+}\right) \,.
\label{lag}
\end{eqnarray}
Here $U =\exp\{2i\vec\tau\cdot\vec \pi/F_\pi\}$
parameterizes the pseudoscalar iso\-triplet of pion fields
$\vec\pi$, $\vec\tau$ are the usual Pauli matrixes,
$\chi=e^{-2\sigma/F_\pi}$ is given in terms of the scalar-isoscalar 
{\em dilaton} field $\sigma$, 
$F_\pi$ is the weak pion decay constant, $e$  is the
parameter of the stabilizing Skyrme term, 
$m_\pi$ is the pion mass, $\varepsilon=16/29$, and
$C_g^*$ is the gluon-condensate parameter.\footnote{The asterisk
indicates that the gluon condensate should be considered as
renormalized one in the nuclear medium. 
Note that in general
$\varepsilon=8 N_f/(11 N_c-2 N_f)$. Furthermore, we assume
$F_\pi=F_\sigma$.} We associate the
dilaton  with  quarkonium. Its mass 
$m_\sigma\approx 600$~MeV is compatible with the well known
attraction in the central NN potential~\cite{Rakhimov}. The skyrmion
is assumed to be located at a position ${\vec R_k}$ from the center of the
nucleus, such that the total spatial vector measured
relative to the
center of the nucleus is given 
as ${\vec x}={\vec R_k}+{\vec r}$, 
where ${\vec r}$ is the distance vector
relative to the origin of the skyrmion.\footnote{This corresponds to
$U=U(\vec x -\vec R_k)=U(\vec r)$ and 
$\chi=\chi(\vec x-\vec R_k)=\chi(\vec r)$ in Eq.\,\re{lag}. 
See  Ref.~\cite{npa002} for further 
details on the geometry of a 
skyrmion (without dilaton) inside a finite nucleus.} 
We have chosen this model as a representative for a larger class of
Skyrme-type models that allow for a qualitative 
description of the central attraction between two nucleons. It is not our
aim to construct a fine-tuned version that produces quantitative fits of
the single nucleon properties or the two-nucleon potentials. Rather, by
keeping
the involved terms to the bare minimum for this case 
(i.e., the non-linear sigma-model kinetic term,
the stabilizing fourth-order derivative term, 
the symmetry-breaking pion mass term, and
the intermediate-attraction generating dilaton term and couplings) 
this model should be general enough to predict qualitatively 
those results which
are generic and also hold 
for more complicated Skyrme-type lagrangians involving
e.g.\ vector-mesons and other hadronic fields.   

The dependence on the 
nuclear density $\rho({\vec x})$
is included in the in-medium coefficients $\alpha_s (\vec x)$ and
$\alpha_p (\vec x)$,
\begin{eqnarray}
\alpha_p(\vec x )&=& 1-\frac{4\pi c_0 \rho(\vec x) /\eta}
{1 +g_0' 4\pi c_0 \rho(\vec x)/\eta}\,,\no\\
\alpha_s(\vec x )&=& 1 - 4\pi \eta b_0 \rho(\vec x)/m_\pi^2\,.
\end{eqnarray}
Here $\eta=1 +m_\pi/m_N \sim 1.14$ is a kinematical factor, and $m_N = 938$~MeV
is the mass
of the nucleon. Moreover, 
$b_0=-0.024\,m_\pi^{-1}$ and $c_0=0.21\,m_\pi^{-3}$ are empirical
parameters that can be taken from the analyses of pionic atoms
and low-energy pion-nucleus scattering data, and
 $g_0'=1/3$ is the Lorentz-Lorenz factor that takes
into account the short-range correlations~\cite{ericson}.

Numerical calculations show that the ground state $B$=1 solution 
of the Euler-Lagrange equations of \re{lag} in  free space  
($\rho=0$, $\alpha_s=\alpha_p=1$) is spherically symmetric.
However, when the soliton is embedded into a finite nucleus, it may alter
its shape, since the spherical  symmetric configuration may not 
correspond to a minimum in energy or mass any longer. In such a way it is 
possible to study the 
modification of the shape of the in-medium skyrmion, namely 
by a minimization procedure of its mass functional.

In deriving the mass functional of the single skyrmion
we follow the scheme presented in
Ref.\,\cite{npa002} by considering an axially
symmetric configuration 
for a  single 
skyrmion located at some distance
from the center of a finite spherical nucleus.
This implies that
the chiral profile function, which parameterizes the modulus of the pion, 
is axially symmetric, i.e.\ $F(\vec r)=F(r,\theta)$,
while the pion direction is governed by the polar-angle profile function 
$\Theta(\theta)$:
\begin{eqnarray}
\vec \pi(\vec r) &=& \dsf{F_\pi}{2}\,
 F(r,\theta) \,{\vec N}(\Theta(\theta),\varphi) \,,\no\\
\vec N&=&\{\sin\Theta(\theta)\cos\varphi,\sin\Theta(\theta)\sin\varphi,
\cos\Theta(\theta)\}\,.
\end{eqnarray}
We further assume that the dilaton field 
also has  an axially
symmetric configuration, i.e.\ 
$\sigma(\vec r)=\sigma(r,\theta)$ and therefore
$\chi(\vec r)=\chi(r,\theta)$. 
For example, the mass functional for single skyrmion 
($B$=1) located at a distance
$R_k$ =$|\vec R_k|$ from the center of the nucleus
and with a spatial vector ${\vec r}_k$ measured
relative to the origin of 
the skyrmion,\footnote{
Note that here $\vec r$ = $\vec r_k$ where $k$ labels the skyrmion. Thus
$\vec x$ = $\vec r_k$+$\vec R_k$ and
$|\vec x|$ = $\sqrt{R_k^2+r_k^2+2 R_k r_k \cos\theta_k}$
= $x(r_k,\theta_k;R_k)$, where $r_k$ and  $\theta_k$ are the modulus and polar
angle of  $\vec r_k$.}
 is given as
\begin{eqnarray}
M(R_k)& =& 2\pi \int\limits_0^\infty d r_k\,r_k^2
\int\limits_{0}^{\pi}d{\theta_k}\sin{\theta_k}
\no\\
&\times& 
\Biggl\{\dsf{F_\pi^2}{8}\phi(F,\Theta;{r_k},{\theta_k})
\,\,\chi^2\,
\alpha_p\left(x(r_k,\theta_k;R_k)\right)
\no\\
&&\mbox{}+\dsf{1}{2e^2}\,\varphi(F,\Theta;{r_k},{\theta_k})\no\\
&&\mbox{}+\dsf{F_\pi^2 m_\pi^2}{4}\left(1-\cos{F}\right)
\,\chi^3\,
\alpha_s\left(x(r_k,\theta_k;R_k)\right)\no\\
&&\mbox{}+\dsf{C_g^*}{24}\psi(\chi)
\label{masstat}
%\left[\chi^4-1+\dsf{4}{\varepsilon}(1-\chi^\varepsilon)\right]\\
+\dsf{F_\pi^2}{8}\left(\dsf{\chi_{\theta_k}^2}{{r_k}^2}
 +\chi_{r_k}^2\right)
\Biggr\}\,,
\end{eqnarray}
\begin{eqnarray}
\phi(F,\Theta,{r_k},{\theta_k})&=&\dsf{F_{\theta_k}^2}{{r_k}^2}
 +F_{r_k}^2+
\dsf{\sin^2F}{{r_k}^2}\left(\dsf{\sin^2\Theta}{\sin^2{\theta_k}}
+\Theta_{\theta_k}^2\right)\,,\no\\
\varphi(F,\Theta;r_k,\theta_k)&=&
\dsf{\sin^2F}{r_k^2}\left[\left(\dsf{\sin^2\Theta}{\sin^2{\theta_k}}
+\Theta_{\theta_k}^2\right)
F_{r_k}^2\right.\no\\
&+&\left.\dsf{\sin^2\Theta}{\sin^2{\theta_k}}
\left(
\dsf{F_{\theta_k}^2}{r_k^2}+
\dsf{\sin^2F}{r_k^2}\Theta_{\theta_k}^2
\right)\right]\,,\no\\
\psi(\chi)&=&
\left[\chi^4-1+\dsf{4}{\varepsilon}(1-\chi^\varepsilon)\right]\no\,.
\end{eqnarray}
Here 
the subscripts of the functions $F_{r_k}$,
$F_{\theta_k}$,
$\Theta_{\theta_k}$,
$\chi_{r_k}$, and
$\chi_{\theta_k}$ denote
the corresponding partial derivatives  (e.g.
$\chi_{\theta_k}\equiv{\partial\chi(r_k,\theta_k )}
 /{\partial{\theta_k}}$ of the function 
$\chi\equiv\chi(r_k, \theta_k )$).

In Ref.\,\cite{npa002} the minimization of the mass
functional~\re{masstat} was performed 
without dilatons, i.e.\ $\sigma(\vec r_k)$=0, 
and the pertinent 
in-medium skyrmion properties were discussed. 
In particular, it was shown that
the medium effects cause a deformation of the nucleon inside finite nuclei.
We follow Ref.\,\cite{npa002} in the minimization procedure for the
functional~\re{masstat}. But in addition 
to the parameterizations of the chiral 
and polar-angle profile functions~\cite{npa002}
\begin{eqnarray}
F&=&2
\arctan\left\{\left(\dsf{r_S^2}{r_k^2}\right)
[1+\gamma_1\cos{\theta_k}+
\gamma_2\cos^2{\theta_k}+\dots]\right\},\no\\
\Theta&=&{\theta_k}+\delta_1\sin 2{\theta_k}+\delta_2\sin 4{\theta_k}
 +\delta_3\sin6{\theta_k}+ \dots\,\,,
\end{eqnarray}
we also introduce 
the following parameterization of the dilaton field:
\begin{equation}
\chi = 1 - \chi_d \exp \left\{-\!\left(\dsf{r_k^2}{r_d^2}\right)\!
 (1+\eta_1 \cos{{\theta_k}}+\eta_2\cos^2{{\theta_k}}+\dots)\right\}.
\label{dilpar}
\end{equation}
Here $r_S$, $\gamma_i$, $\delta_i$, $\chi_d$, $r_d$ 
and $\eta_i$ are variational parameters.
As a result of the
minimization procedure one gets a set of variational parameters
corresponding to specified values of  the distance  $R_k$ 
and the other
input parameters defined in subsection~\ref{inputpar} and 
section~\ref{results}.

In this context we should reiterate that we did not choose the model such that
it
reproduces in a fine-tuned way  
quantitatively 
the single nucleon properties or  the two-nucleon interaction in free space,
but that it rather serves as  generic representative of Skyrme-type models  by 
describing  qualitatively the {\em modifications} when the single skyrmion or
the two-skyrmion system is embedded in a {\em finite} nucleus.

\subsection{Two nucleons in a baryon-rich environment and the residual
NN interaction}
\begin{figure}[hbt]
\begin{center}
\epsfxsize=7cm
\epsffile{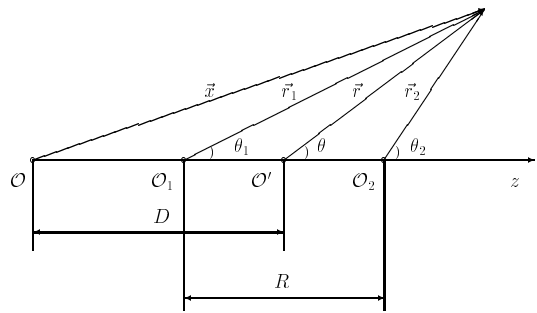}
\end{center}
\vskip 0.03cm
\caption{
\label{fig1}
The axially symmetric
two-skyrmion system in a finite spherical nucleus. Here ${\cal O}$ is the
center of the nucleus, ${\cal O}'$ is the geometrical center of
the two-skyrmion system, ${\cal O}_1$ and ${\cal O}_2$ are the
centers of the first and second skyrmions, respectively. $D$ is
the distance between the center of the nucleus and the one of the two-skyrmion
system, and  $R$ is the distance between skyrmions. 
%The origin of the
%coordinate system coincides with the 
%geometrical center of the two-skyrmion system ${\cal O}'$.
}
\end{figure}

We now turn  to  the system of two interacting in-medium 
nucleons,
e.g.\ to the  modification of NN potential $V_{NN}$ inside nuclei.
As mentioned in the introduction, we only consider the modification of the
NN potential in the {\em radial} direction, since the modification in the
{\em azimuthal} direction is expected to be subleading.  
The geometry of this axially symmetric case is  presented in Fig.\ref{fig1}.
%%%%%%%%%%%%%%%%%%%% figure 1

Unfortunately, in any version of the Skyrme lagrangian,
it is very difficult to derive the binding
energies as well as the form factors of even light nuclei
directly from the nucleon-nucleon potential $V_{NN}$. 
For simplicity, in setting up  our formalism, we assume that
\begin{itemize}
\item
each nucleon is bound inside the finite nucleus
by a phenomenological averaged potential normally used in shell
models in MFA;
\item
the nucleon distribution, i.e.\ the nuclear density, is given
by a phenomenological formula;
\item
 the primary two-nucleon potential in free space
is described by the lagrangian of the  Skyrme model~\re{lag}
with $\alpha_p=\alpha_s=1$.
\end{itemize}
Although for practical calculations of nuclear properties in  MFA
the explicit expression of $V_{NN}$ is not needed, this does not
mean that the nucleons do not interact with each other. In fact, it
is believed that the sum 
of all two-body -- as well as possible three-body -- 
interactions can be reexpressed as  an effective averaged potential,
centered at the origin of the nucleus. Due to the MFA
ideology, every nucleon behaves as if it ``feels'' only this
effective potential (spherical-well potential, oscillator,
etc.). Yet residual interactions between nucleons may
still exist. These interactions ought to be important for the description of
deformations and stabilities of the nuclei. 

In the following we will
study the residual part of NN interactions (in radial direction) 
embedded in the surrounding nuclear
environment. We assume that the primary two-nucleon potential in
free space is given by the Skyrme model, including dilatons,
as described by the lagrangian~\re{lag} with density $\rho=0$.  
We then ``plug'' a pair
of nucleons into a finite nucleus, or more exactly, we consider
them as part of the nucleus.
The point is that
each of the  nucleons can be deformed for the
following two reasons, which
affect the in-medium potential
as well:
\begin{itemize}
\item[1)]
the presence of
the other nucleon or, in other words, the presence of the NN interaction;
\item[2)]
the presence of
the remaining $A\!-\!2$ nucleons or, in other words, the
presence of medium effects in the nucleus.
\end{itemize}
The first case was investigated in Refs.~\cite{Otofujii,Rakhimov2},
where the modifications of the nucleons and $V_{NN}$ properties in free
space were  recorded under possible deformations of the nucleons during
their mutual approach. For that purpose the authors introduced a couple of 
deformation
parameters and obtained a modified central potential $V_{NN}^c$
by minimizing it at each internucleon distance $R$. The second
case, namely the effects of the nuclear medium, 
is the subject of this article. 
In principal, we
could 
have followed the procedure of Refs.~\cite{Otofujii,Rakhimov2},
which would be:
\begin{itemize}
\item[{1)}]
 to introduce deformation parameters in
lagrangian~\re{lag}
using an axially symmetric ansatz;
\item[{2)}]
to construct $V_{NN}^c$ from the lagrangian~\re{lag} in the product
approximation;
\item[{3)}]
 to determine the deformation parameters by minimizing
$V_{NN}^c$ at each $R$.
\end{itemize}
However such a procedure would be unnecessarily complicated, 
especially for the case of
lagrangian~\re{lag} with dilatons. Moreover, it that case one 
would have to fight complications  
resulting from double
counting the deformation effects caused by NN interactions and 
by medium modifications, respectively.
Instead, for simplicity and also for isolating the medium modifications from
the vacuum NN interaction effects, 
we minimize the mass functional~\re{masstat}
for each of the two in-medium skyrmions with respect to the corresponding
deformation parameters,
and then we insert  these deformed skyrmion 
profiles into the product ansatz in order to generate
the NN central
potential.

In this way, by inserting two skyrmions in the nuclear background of the
$A\!-\!2$ 
remaining nucleons, one can define a static skyrmion-skyrmion residual
potential
\begin{eqnarray}
&&V_{SS}(\underbrace{\frac{1}{2}|\vec{R_1}+\vec{R_2}|}_{\equiv D},
\underbrace{|\vec{R_2}-\vec{R_1}|}_{\equiv R})\no\\
&&\quad
=M[U(\underbrace{\vec x - \vec R_1}_{\equiv \vec r_1})
U(\underbrace{\vec x - \vec R_2}_{\equiv \vec r_2}),
\chi(\vec r_1)\chi(\vec r_2);
\rho_{A-2}(|\vec x|)]
\no\\
&&\qquad
-2M^{free}_{B=1}-\delta V^{MF}_{B=2}(D,R)\,.
\label{vnn0}
\end{eqnarray}
Here $M[\prod_{i=1}^2
U(\vec x-{\vec R_i}),\chi(\vec x - \vec R_i);\rho_{A-2}(|\vec x|)]$
is the static mass functional of a
skyrmion-product ansatz of baryon number $B=2$
with the skyrmions centered (axial-symmetrically) at
$\vec{R_i}\equiv {\vec{{\cal O}{\cal O}_i}}$ 
(with ${\cal O}$ and ${\cal O}_i$
as defined in Fig.\ref{fig1})
in the renormalized
background density $\rho_{A-2}(|\vec x|)$.
Thus the potential~\re{vnn0} is defined
in analogy with the well known product-ansatz prescription of the dilaton-free
case 
\begin{equation}
  U_{B=2}\left(\vec x- \vec D ,\vec R\right)
 =U(\vec x-\vec R_1)U(\vec x-\vec R_2)\,,
\end{equation} 
where $\vec D\equiv(\vec R_1+\vec R_2)/2$ and $\vec R\equiv \vec R_2-\vec R_1$.
The remaining quantity $\delta V^{MF}_{B=2}$ 
denotes the contribution of the $B$=2 system to the mean-field
potential.
It is defined as
\begin{equation}
\delta V^{MF}_{B=2}=
\!
\sum_{i=1}^2\left(
M[U(\vec r_i),\chi(\vec r_i);\rho_{A-2}(|\vec x|)]
-M^{free}_{B=1}\right).
\label{MFP}
\end{equation}
Note that the  corresponding quantity $\delta M_{S}=-\delta
V_{B=1}^{MF}$ can be interpreted 
as  the part of the energy of skyrmion  that is 
transformed to the mean-field energy of pions inside the nucleus and 
to the interaction energy of the skyrmion
with the background field. It is this background pionic field that generates
the attractive mean field potential binding nucleons to the
nucleus.
Clearly the potential~\re{vnn0} becomes the free
NN potential when the background density
of the remaining nucleons goes to
zero\footnote{If the background field vanishes, 
$\rho_{A-2}\rightarrow 0$,
the mean-field contribution also vanishes, 
$\delta V^{MF}_{B=2}\rightarrow 0$.}.

Finally, inserting~\re{MFP} into~\re{vnn0}, one gets the following form of
the skyrmion-skyrmion residual potential
\begin{eqnarray}
V_{SS}(D,R)
&=&M[U(\vec r_1)U(\vec r_2),\chi(\vec r_1)\chi(\vec r_2);
\rho_{A-2}(|\vec x|)]
\no\\
&&
-M[U(\vec r_1),\chi(\vec r_1);\rho_{A-2}(|\vec x|)]
\no\\
&&
-M[U(\vec r_2),\chi(\vec r_2);\rho_{A-2}(|\vec x|)]\,.
\label{vcc}
\end{eqnarray}
From the comparison of Eqs.\,\re{vnn0}, \re{MFP}, and \re{vcc} it should 
become 
clear that the potential~\re{vcc} is simply the difference
between the separation energy of a pair of skyrmions ($B$=2) inside the nucleus
and the sum of the separation energies of each individual 
skyrmion ($B$=1) in the same nucleus.
This is well known in nuclear physics as a
major contribution to the residual NN potential in the nuclear medium.
Our aim is the evaluation of the spin-isospin independent part of 
this potential. 
potential. Note that in this case  
the skyrmion-skyrmion interaction and the nucleon-nucleon interaction
coincide.

As mentioned before, we only consider the geometry 
when the center of the (spherical) 
nucleus and centers of the two skyrmions fall
onto the same axis,
which is denoted by $z$ as illustrated in Fig.\ref{fig1}.
Then the 
$V_{NN}$ potential depends only on
two variables:
1) the distance between the centers of the two skyrmions, $R$; 2) 
the distance between
the geometrical center
of the two-skyrmion system and the center of the nucleus, $D$.
The spatial vector $\vec r$ is measured relative to the two-skyrmion center, 
such that here $\vec x = \vec D +\vec r$.

To isolate  the spin-isospin independent part of the  
NN potential one can use the
standard projection procedure~\cite{Nyman},
which corresponds to the following form in free space
\begin{eqnarray}
V_{NN}(R,C)&=&V_{NN}^c
 +\dsf{1}{2}\Tr(C\sigma_iC^+\sigma_j)
\Bigl[(3\hat R_i\hat R_j\!-\!\delta_{ij})\no\\
&&\qquad\mbox{}\times 
V_{NN}^T(R)+
\delta_{ij}V_{NN}^{\sigma\tau}(R)\Bigr]\,,
\label{vnn-last}
\end{eqnarray}
where  $C=u_0+i\vec\tau\cdot{\vec u}$ (with ${\vec u}^2+u_0^2=1$) 
is the
relative orientation matrix of the
nucleons in internal space, $\sigma_{i,j}$ are the Pauli matrixes,
 and $\hat R$ is the unit vector along the line joining
the centers of the two nucleons.
Here $V_{NN}^c$, $V_{NN}^T$ and $V_{NN}^{\sigma\tau}$ are
the
central, 
tensor and spin-spin parts of the NN potential, respectively.

Finally, taking into account the projection procedure~\cite{Nyman} and
using the expression for skyrmion-skyrmion potential~\re{vcc}, we
predict 
the following expression for the spin-isospin--independent part of the
residual NN potential (in radial direction) inside a finite spherical nucleus:
\begin{equation}
V_{NN}^{c,A}(D,R)
=
2\pi\int\limits_0^\infty\!\! d r\, r^2\!\!
\int\limits_0^\pi\!\! d\theta\sin\theta
\bigl\{ V_2^c \!+\! V_4^c\!+\! V_{\chi SB}^c
\!+\!  V_\sigma^c\bigr\}%\,,
\end{equation}
with
\begin{eqnarray*}
 V_2^c&=&\dsf{F_\pi^2}{8}
 \alpha_{p,A-2}\left(x(r,\theta;D)\right)
 \bigl[\chi_1^2\chi_2^2(\phi_1+\phi_2)\no\\
&&\qquad\qquad\mbox{}
-(\chi_1^2\phi_1
+\chi_2^2\phi_2)\bigr],\no\\
 V_4^c
&=&\dsf{1}{3e^2}(\phi_1\phi_2-\Phi)\,,\no\\
%%%%+\dsf{1}{2}(\varphi_1+\varphi_2)\,,\no\\
 V_{\chi SB}^c
&=&
\dsf{m_\pi^2 F_\pi^2}{4}
\alpha_{s,A-2}\left(x(r,\theta;D)\right)
\bigl[\chi_1^3\chi_2^3(1\!-\!\cos F_1\cos F_2)\no\\
&&\mbox{}-\chi_1^3(1-\cos F_1)-\chi_2^3(1-\cos F_2)\bigr]\,,\no\\
 V_\sigma^c
&=&\dsf{C_{g,A-2}^*}{24}\Bigl[\psi(\chi_1\cdot\chi_2)
-\psi(\chi_1)-\psi(\chi_2)\Bigr]\no\\
&+&\dsf{F_\pi^2}{8}\left(\chi_1^2-1\right)
\left(\dsf{\chi_{2,\theta_2}^2}{r_2^2}
+\chi_{2,r_2}^2\right)\no\\
&+&\dsf{F_\pi^2}{8}\left(\chi_2^2-1\right)
\left(\dsf{\chi_{1,\theta_1}^2}{r_1^2}
+\chi_{1,r_1}^2\right)\no\\
&+&\dsf{F_\pi^2}{4}\chi_1\chi_2 \left[\sin(\theta_1\!-\!\theta_2)\!
\left(\!\chi_{1,r_1}\dsf{\chi_{2,\theta_2}}{r_2}-
\dsf{\chi_{1,\theta_1}}{r_1}\chi_{2,r_2}\!\right)\right.\no\\
&& \mbox{}+\left.\cos(\theta_1\!-\!\theta_2)
\left(\chi_{1,r_1}\chi_{2,r_2}
+\dsf{\chi_{1,\theta_1}\chi_{2,\theta_2}}
{r_1 r_2}\right)\right]\,,\no\\
\Phi
&=&\left[\sin(\theta_1-\theta_2)
\left(F_{1,r_1}\dsf{F_{2,\theta_2}}{r_2}-
\dsf{F_{1,\theta_1}}{r_1}F_{2,r_2}\right)\right.\no\\
&&\mbox{}+\left.\cos(\theta_1-\theta_2)
\left(F_{1,r_1}F_{2,r_2}+\dsf{F_{1,\theta_1}F_{2,\theta_2}}
{r_1 r_2}\right)\right]^2\no\\
&+&\dsf{\sin^2{F_1}}{r_1^2}\Biggl(\cos(\theta_1-\theta_2)
\dsf{F_{2,\theta_2}}{r_2}\no\\
&&\mbox{}-\sin(\theta_1-\theta_2)F_{2,r_2}\Biggr)^2
\Theta_{1,\theta_1}^2\no\\
&+&\dsf{\sin^2{F_2}}{r_2^2}\Biggl(\cos(\theta_1-\theta_2)
\dsf{F_{1,\theta_1}}{r_1}\no\\
&&\mbox{}+\sin(\theta_1-\theta_2)F_{1,r_1}\Biggr)^2
\Theta_{2,\theta_2}^2\no\\
&+&\dsf{\sin^2{F_1}}{r_1^2}\dsf{\sin^2{F_2}}{r_2^2}
\Biggl(\dsf{\sin^2{\Theta_1}}{\sin^2\theta_1}
\dsf{\sin^2{\Theta_2}}{\sin^2\theta_2}\no\\
&&\mbox{}+\cos^2(\theta_1-\theta_2)
\Theta_{1,\theta_1}^2\Theta_{2,\theta_2}^2\Biggr)\,.
\end{eqnarray*}
Here
$F_i\equiv F_{i}(r_i,\theta_{i})$,
$\chi_i\equiv \chi_{i}(r_i,\theta_{i})$,
$\Theta_i\equiv \Theta_{i}(\theta_{i})$ with $i=1,2$ 
are the chiral, dilaton, and polar-angle 
profile functions of skyrmion 1 or 2,
respectively, with 
$r_i=r_i(r,\theta;R)=\sqrt{(R^2/4)+r^2\pm R r\cos\theta}$,
$\theta_i=\theta_i(r,\theta;R)=\arcsin(\sin\theta\,r/r_i(r,\theta;R))$,  
$r$, $\theta$ and $x$ =
$x(r,\theta;D)$ = $\sqrt{D^2+r^2+2D r \cos\theta}$ 
as defined in Fig.\,\ref{fig1}. Moreover, we abbreviate the partial
derivative of the above defined functions 
$f_i\in\{F_i,\chi_i,\Theta_i\}$ as
$f_{i,y}\equiv {\partial f_i}/{\partial y}$. 
The functions $\phi_i$
%%%and $\varphi_i$
are defined as
$\phi_i\equiv \phi(F_i,\Theta_i;r_i,\theta_i)$ with 
$\phi$   (and $\psi$) as in Eq.\,\re{masstat}.
%%%$\varphi_i=\varphi(F_i,\Theta_i;r_i,\theta_i)$.

\subsection{Density dependence and input parameters}
\label{inputpar}

We shall use a phenomenological parameterization of the density of
a spherical nucleus as presented in Ref.\,\cite{Akhiezer} and therefore write
the renormalized density of the $A-B$ nuclear background environment
as
\begin{eqnarray}
%\begin{array}{lll}
\rho_{A-B}(x)&=&\left(\dsf{A-B}{A}\right)
\dsf{2}{\pi^{3/2}r_0^3}\left[1+\dsf{A-2}{3}
\left(\dsf{x^2}{r_0^2}\right)\right]\no\\
&&\mbox{}\times
\exp\left\{-\dsf{x^2}{r_0^2}\right\} \qquad\qquad\qquad \mbox{if } A<20,\no\\
\rho_{A-B}(x)&=&\left(\dsf{A-B}{A}\right) %\\
%&&\mbox{}\times
\dsf{\rho_0}{1+\exp\{(x-R')/a\}} \no\\
&&\qquad\qquad\qquad\qquad\qquad\qquad\qquad \mbox{if } A\ge 40.\no
%\end{array}
\end{eqnarray}
where
 $r_0=1.635$\,fm as for $^{12}$C. Furthermore, 
$a=0.58$\,fm, $R'=1.2A^{1/3}$\,fm, and
$\rho_0=0.5 m_\pi^3$ is the normal nuclear matter density.

The input parameters of the skyrmion sector
are chosen as $F_\pi=$186~MeV and $e=2\pi$. Although this standard
set of input parameters
gives an overestimated
value for the
mass of nucleon even in free space, we shall not optimize it,
since we are only interested in the influence
of the medium on the NN potential.
The sole
input parameter in the dilaton sector is the gluon condensate parameter
$C_g^*$. Its 
in-medium renormalization has not been clarified in the literature. In the
present model this quantity can be expressed in terms of the in-medium
sigma-meson mass $m_\sigma^*$ as~\cite{Andrianov88,Musakhanov}:
\begin{equation}
C_{g,A-B}^*=\dsf{3F_\pi^2m_{\sigma,A-B}^{*2}}{2(4-\varepsilon)}.
\end{equation}
Unfortunately, there is also no information about the renormalization of
$m_\sigma$ within the present approach. Various 
approaches~\cite{Brown2,Meissner89,Saito} show that $m_\sigma^*$ has 
a linear density dependence.
In view of this, we shall use the parameterization
\begin{equation}
m_{\sigma,A-B}^*(x)=\left(1-0.12\,\dsf{\rho_{A-B}(x)}{\rho_0}\right)m_\sigma
\end{equation}
that was obtained in Ref.\,\cite{Saito} within the
quark-meson coupling model.\footnote{
Since in finite nuclei the density is coordinate-dependent,
$m_\sigma^*$ will also acquire a coordinate dependence.}
The free-space mass of the sigma meson
is taken as $m_\sigma=550$~MeV, which corresponds to a free-space 
value of $C_g=(260~\mbox{MeV})^4$.

\section{Results and discussion}
\label{results}

The results
of the minimization procedure for a single skyrmion ($B=1$)
in the nucleus with an  $A-2$ background density is
presented in Table\,\ref{tab1}.
\begin{table}[htb]
\begin{center}
\begin{tabular}{crrrrr}\hline\hline
\multicolumn{6}{c}{$^{12}$C}
\\\hline
 $R_k$[fm] &  0   & $\pm$0.649 & $\pm$1.298 & $\pm$1.947 & $\pm$2.595 \\
 $r_S$[fm] &0.512 & 0.509      & 0.495      & 0.467      & 0.435  \\
 $\gamma_1$&  0   &$\pm$0.020  &$\mp$0.034  &$\mp$0.105  &$\mp$0.107 \\
 $\gamma_2$&  0   & -0.028     & -0.047     &  0.001     &  0.037 \\
 $10\delta_1$&0   & -0.056     & -0.097     & -0.018     &  0.047 \\
 $10\delta_2$&0   &   0        & -0.002     & -0.003     & -0.001 \\
 $\chi_d$    
           & 0.634&  0.629     &  0.592    &  0.655     &  0.766 \\
 $r_d$[fm]
     &      0.843 &  0.841     &  0.835     & 0.808      &  0.776 \\
 $\eta_1$    &  0 &  $\pm$0.055&$\pm$0.008  &$\mp$0.060  &$\mp$0.054\\
 $\eta_2$    &  0 & -0.015     &  -0.032    &      0     &   0.017\\
\hline\hline
\multicolumn{6}{c}{$^{40}$Ca}\\\hline
 $R_k$[fm]   & 0    &$\pm$1.705 &$\pm$3.409 &$\pm$5.114 &$\pm$6.819 \\
 $r_S$[fm]   & 0.597&0.576      &0.477      &0.407      &0.398 \\
 $  \gamma_1$& 0    &$\mp$0.053 &$\mp$0.145 &$\mp$0.035 &$\mp$0.002 \\
 $  \gamma_2$& 0    &-0.011     & 0.030     & 0.017     & 0.001  \\
 $10\delta_1$& 0    &-0.026     & 0.021     & 0.027     & 0.002 \\
 $10\delta_2$& 0    & 0         &-0.001     & 0         & 0     \\
 $\chi_d     $      
             & 0.397& 0.423     & 0.650     & 0.854     & 0.876 \\
 $r_d$[fm]
             & 0.983& 0.951     & 0.818     & 0.749     & 0.741 \\
 $\eta_1$    & 0    &$\mp$0.020 & $\mp$0.076&$\mp$0.010 & $\mp$0.001 \\
 $\eta_2$    & 0    & -0.005    & 0.014     & 0.004     &   0\\
\hline\hline
\end{tabular}
\end{center}
\caption{
\label{tab1}
Deformation parameters at various separations
$R_k$ of the center of a single skyrmion from the center of the nucleus. 
The parameters $r_S$, $\gamma_i$, $\delta_i$
are for the chiral sector, while $\chi_d$, $r_d$, $\eta_i$ are for the
dilaton sector. They are obtained by a direct minimization
of the mass functional in~\re{masstat}. Note that all coefficients
$\gamma_i$, $\delta_i$, $\eta_i$ for $i\ge 3$ are negligible and
are not presented here. Non-zero free-space values of some parameters are:
$r_S =0.398$\,fm, $\chi_d= 0.877$, $r_d=0.740$\,fm.}
\end{table}
We have not
listed all variational parameters, as some of them
are rather small, namely
$\gamma_i$, $\delta_i$, $\eta_i$ for $i\ge 3$. 
These parameters, which describe higher multipole deviations 
of the skyrmion shape from the
spherical form, can be neglected without loss of accuracy.
A discussion about the parameters of the chiral sector can be found in
Ref.\,\cite{npa002}, whereas the ones of the dilaton sector are new. 
Both sets  are consistent with the fact that the nucleon inside the nucleus
acquires an intrinsic quadrupole moment unless the nucleon is located directly
at the center.
Furthermore note that the increase of the radial parameter $r_S$ of the chiral
sector and of the radial parameter $r_d$ of the dilaton sector with increasing
density is in 
qualitative agreement with the overall picture of the swelling of the 
nucleon inside the nucleus. The behavior of the depth of the dilaton field 
parameterized by $\chi_d$  
is more involved, but at least for the case of the heavier nucleus 
the dilaton region
of the in-medium nucleon becomes
shallower with increasing density. At first sight this is counter-intuitive.
However, note that because of the swelling of the nucleon the total 
region where the ``effective pion decay parameter'' 
$\chi F_\pi$ is reduced relatively to
its vacuum value $F_\pi$ has increased nevertheless.

\begin{figure}[ht]
\begin{center}

   \epsfxsize=7cm
\epsffile{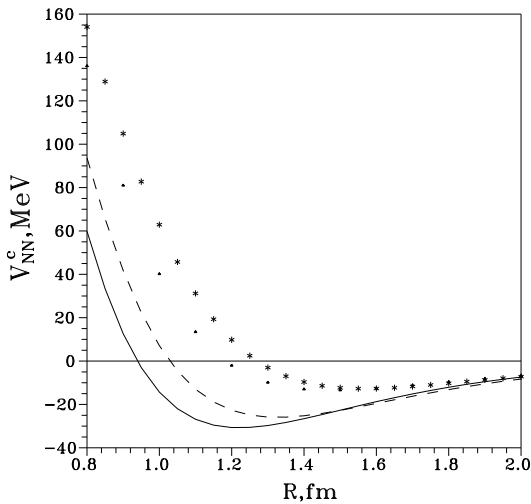}
\end{center}
%\vskip -8cm
\caption{
\label{vcc12}
Spin-isospin-independent central part (in radial direction) 
of the residual NN
potential in $^{12}$C. The solid line corresponds to the potential in
the free case, the dashed line corresponds to $D=3$~fm, the dotted line
corresponds to $D=2$~fm, and the stars correspond to  $D=0$~fm. }
\end{figure}

\begin{figure}[ht]
\begin{center}
   \epsfxsize=7cm
\epsffile{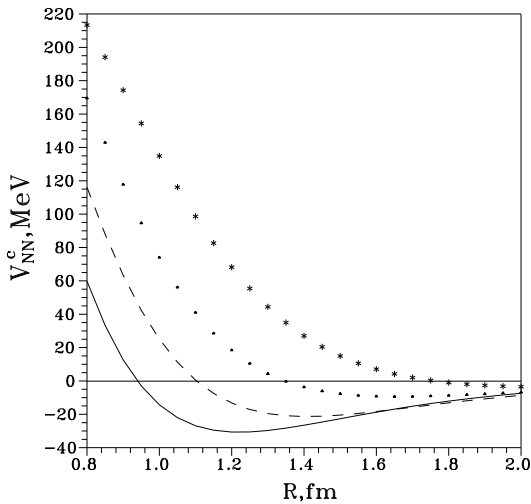}
\end{center}
%\vskip -8cm
\caption{
\label{vcca40}
Spin-isospin-independent central part (in radial direction) 
of the residual NN
potential in $^{40}$Ca. The solid line corresponds to the potential
in the free case, the dashed line corresponds to $D=4$~fm, the dotted line
corresponds to $D=3$~fm, and the stars correspond to $D=0$~fm. }
\end{figure}

The results of the calculations for the residual NN interactions 
are summarized in Figs.~\ref{vcc12} and~\ref{vcca40}.
%%%%%%%%%%%%%%%%%%% fig2
%%%%%%%%%%%%%%%%%%%%%%%% figure 3
The spin-isospin-independent central part $V_{NN}^c$ of the
residual NN potential in $^{12}$C is shown in Fig.\,\ref{vcc12}.
Let us emphasize again that our model has not been fine-tuned 
to reproduce the spin-isospin independent NN potential in free space 
quantitatively. But the qualitative changes are evident:
One can see that short-range repulsive contribution of the potential is
increased and the intermediate attraction is decreased in the
nuclear medium. The same behaviour of $V_{NN}^c$ is observed for
skyrmions embedded in $^{40}$Ca,  as shown in
Fig.\,\ref{vcca40}. 

It is known that 
the residual NN potential becomes repulsive when
neighboring nucleons overlap. It  is furthermore known that the
sizes of the nucleons increase in the medium, see also the parameters 
$r_S$ and  $r_d$ of
Table~\ref{tab1}. Therefore 
the overlap-regions increase with
density. This leads to a build-up of the repulsive part of the
potential at high densities, i.e.\ when the two-nucleon system 
is near the center of the nucleus.

From previous studies~\cite{npa002} we know that the nucleon mass is the
smallest near the center of nucleus and increases 
when the nucleon is moved to
the surface. Due to this phenomenon, nucleons should actually collapse to the
center of the nucleus. In other words, due to the attractive mean-field
potential, nucleons should move to the center. On the other hand, the nucleon
concentration near the center of the nucleus does, as shown in
Figs.\,\ref{vcc12} and \ref{vcca40}, increase the repulsive residual
potentials between nucleons such that nucleons are expelled from the center.
Consequently, an equilibrium state arises and therefore saturation of the
nuclear matter density results. 
In this case nucleons stop their radial motion
towards the center of the nucleus, but their motion in a shell with given
radius continues. A further inclusion of the Pauli mechanism in this picture
could finally provide for the shell description of finite nuclei.

Su {\em et al.} obtained similar
results to ours in the framework
of a chiral $\sigma$--$\omega$ model under
the imaginary--time Green's function method~\cite{Su94}. 
They found that the potential well of the nucleon-nucleon interaction
becomes shallow as the temperature increases.
The same behaviour of the NN potential was seen in 
quantum hadrodynamics studies~\cite{Gao95} and 
in the framework of TFD~\cite{sgao}. The latter established, 
in addition, 
the alternate roles of density and temperature effects.

\section{Summary and outlook}

We have considered the central part of the residual NN
potential inside finite nuclei in the radial direction. 
In the interior region of the heavy
nucleus the residual NN potential is strongly repulsive and
compensates the mean-field attractive potential in a such way that
the nuclear density saturates.  These results are consistent  
with other studies  applying different 
methods~\cite{sgao,Su94,Gao95}. 
However, one should remember 
that the Skyrme-type models are the only models that 
are formulated in terms of hadronic (non-quarkish) degrees of freedom 
{\em and} 
that can simultaneously describe
both single-baryon as well as multi-baryon properties. In the framework of
such  a class of models our results are indeed new.

It
would be interesting to investigate the role of the residual
interactions in the formation of nuclear matter itself, i.e. to
study their role in the formation of the mean-field potential. This
requires the additional study of the in-medium NN potential in
connection with nuclear matter properties. For example, for
some relative orientations of nucleons in the internal space (see
Eq.~\re{vnn-last}) the tensor part of the potential $V_{NN}^T$
makes a negative contribution to the total NN potential. Such an
attraction  turned out to be sufficient for the construction of 
crystalline~\cite{Klebanov85} or condensate~\cite{Dyakonov88}
states of nuclear matter. Note, however, that medium modifications of
the NN potential were neglected. Previous
calculations~\cite{prc001} showed that the tensor part of the
residual potential does decrease in the nuclear medium. 
Thus  the Skyrme model allows to combine
the tensor and the central potentials into a full
potential which is  shallow due to the medium influence.
The question is whether the changes 
of the in-medium NN potential
are sufficient to 
induce a breakdown of the crystalline
structure which seems not to exist in nature. Of course 
the medium modifications of the NN potential might only be one of 
many agents that could induce this breakdown.
For instance, quantum fluctuations~\cite{Dyakonov88,Walhout88} 
and Fermi-motion effects should be of importance~\cite{Cohen89}.
It is well known that the nuclear matter binding results
from a strong cancellation of an attractive (binding) potential
term and a repulsive kinetic term.
It remains to be seen whether the  condensed state of 
nuclear matter at ordinary densities can be achieved within the in-medium 
modified Skyrme model.

This research is part of the EU Integrated Infra\-structure Initiative
Hadron Physics Project under contract number RII3-CT-2004-506078 and
is supported in  part by DFG (SFB/TR 16) and by the Forschungszentrum J\"ulich
under contract No.\,41445400 (COSY-067).
U.T.Y. would like to thank A.K.~Nasirov for useful
correspondence and acknowledges support
by Pusan National University 
(Post-Doc 2004 program).

\end{document}